\documentclass[amsmath,preprintnumbers,10pt,nofootinbib,prl,twocolumn]{revtex4-1}
\usepackage{amsbsy}
\usepackage{amssymb}
\usepackage{graphicx}
\usepackage{color}
\usepackage{subfigure}
\usepackage{physics}
\usepackage{soul}
\usepackage{color}
\usepackage{bm}
\usepackage[normalem]{ulem}

\usepackage{verbatim}
\usepackage{natbib}
\begin{document}
\title{Dissipation induced transitions in two dimensional elastic membranes}
\author{Michael Nguyen$^{1,2}$, Suriyanarayanan Vaikuntanathan$^{1,2}$} 
\affiliation{$^1$The James Franck Institute, The University of Chicago, Chicago, IL,}
\affiliation{$^2$ Department of Chemistry, The University of Chicago, Chicago, IL.}
\begin{abstract}
Stochastic thermodynamics provides a useful set of tools to analyze and constrain the behavior of far from equilibrium systems. In this paper, we report an application of ideas from stochastic thermodynamics to the problem of membrane growth. Non-equilibrium forcing of the membrane can cause it to buckle and undergo a morphological transformation. We show how ideas from stochastic thermodynamics, in particular the recently derived thermodynamic uncertainty relations, can be used to phenomenologically describe and constrain the parameters required to excite morphological changes during a non-equilibrium growth process.

\end{abstract}
\maketitle

\noindent{\it Introduction:} Non-equilibrium forcing can be used to uncover new strategies for self-assembly and organization~\cite{Battle604,Lan2012,Mehta2012,Whitelam2014}. In biophysical contexts, it has been established that non-equilibrium forces play a crucial role in suppressing rogue fluctuations and enhancing fidelity of molecular recognition~\cite{Hopfield1974,Mehta2012,Murugan2012,Murugan2016,Vaikunt2017}, support robust oscillations crucial for the maintenance of circadian rhythms~\cite{Barato2017}, and drive sensory adaptation processes~\cite{Lan2012,Mehta2012}. Non-equilibrium forces also play an important role in modulating cell shape and cell membrane fluctuations~\cite{McMahon2005, Turlier2016}. For instance, local changes in surface tension or lateral pressure due to a spontaneous assembly of membrane proteins have been known to induce instabilities in membrane fluctuations~\cite{Stachowiak2012, Chen2016,Rangamani2014,Leibler1986}. Such instabilities have been implicated as important precursors during cell division~\cite{McMahon2005}. Non-equilibrium fluctuations are also important in cases where the cell membrane interacts with growing actin filaments. The important role played by such interactions in regulating the organization of the membrane has been well established ~\cite{Gowrishankar2012,Weichsel2016}.
\begin{figure}[tbb]
\centering
\includegraphics[scale=0.35]{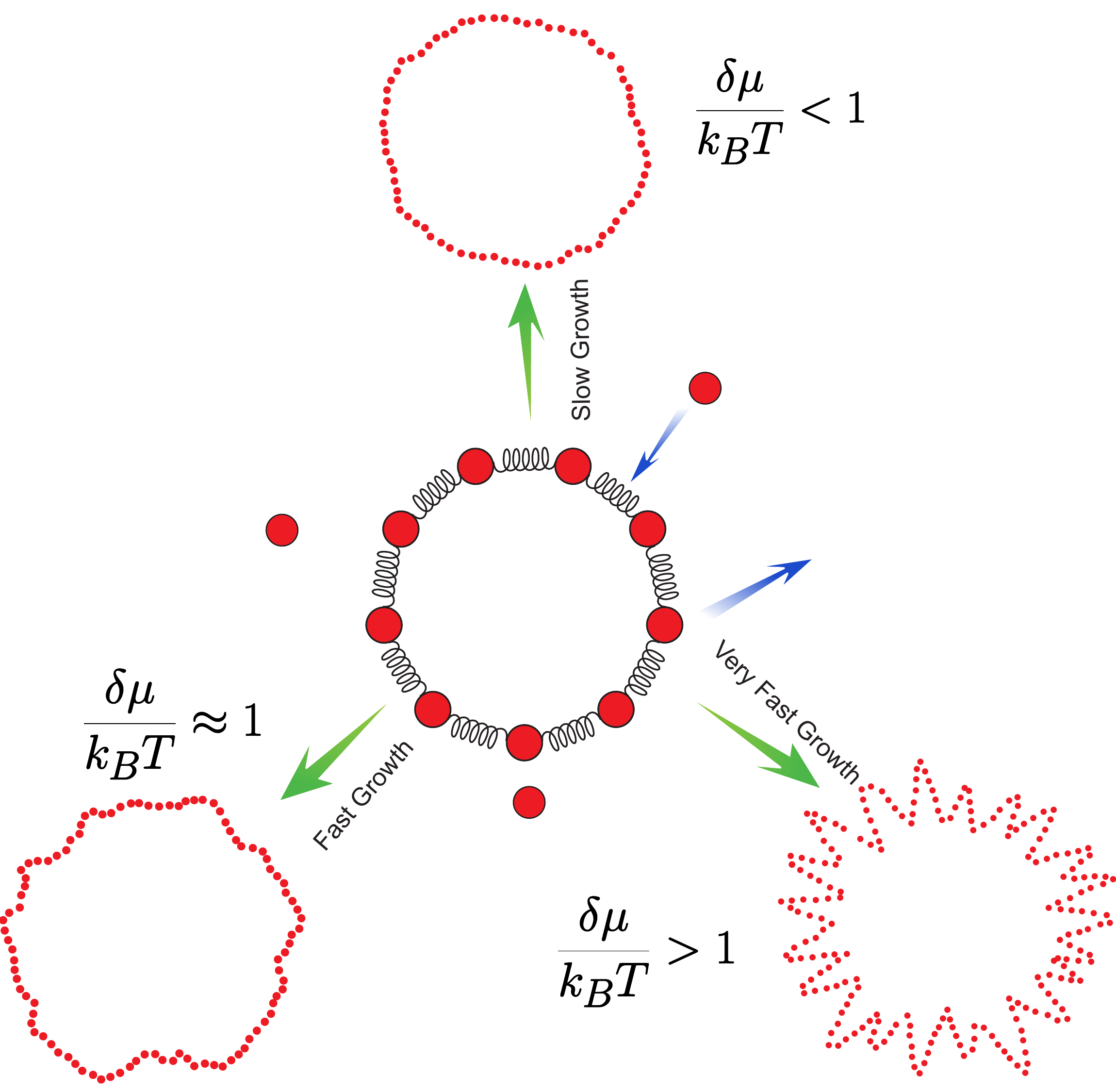}
\caption{Schematic of the growing assembly. When the assembly grows slowly, its shape remains circular (top figure). As the assembly grows faster, its shape becomes more distorted (lower left) and ultimately it buckles into a star-shaped morphology (lower right). We have included movies in the SI showing this transformation.} \label{fig:phases}
\end{figure}

However, unlike the behavior and characteristics of equilibrium systems, general principles governing self-assembly and organization away from equilibrium remain to be discovered. In this letter, we use the framework of stochastic thermodynamics to investigate non-equilibrium growth and morphological changes~\cite{Drasdo2000, Ramaswamy2000,Rao2001,Solon2006,Hannezo2011,Fisher1989,Fisher1989a,Rudnick1991,Rajesh2008,Nelson2009,Mahadevan2019,Wolde2019} in a model elastic membrane (Fig.~\ref{fig:phases}). Our model consists of two-dimensional particles connected by elastic springs in a ring like geometry (Fig~\ref{fig:phases})~\cite{Fisher1989,Fisher1989a,Rudnick1991,Rajesh2008,Nelson2009}. The ring assembly is allowed to exchange particles with a reservoir. The chemical potential of the reservoir controls the growth rate of the ring assembly and sets the non-equilibrium driving force in this system. This elastic model is adapted from an equilibrium model first introduced by Leibler and coworkers in Ref~\cite{Fisher1989}. As demonstrated in Refs~\cite{Fisher1989,Fisher1989a,Rudnick1991,Rajesh2008,Nelson2009}, despite their apparent simplicity, this class of elastic models possess many features~\cite{W.Helfrich1973,Fisher1989,Fisher1989a,Rudnick1991,Ramaswamy2000,Rao2001,Solon2006,Rajesh2008,Hannezo2011,Loubet2012,Nelson2009} characteristic of three-dimensional membranes and can be used to obtain insights into how morphological changes in such systems can be excited under a non-equilibrium driving force.

Indeed, as we describe below, our numerical analysis shows that the effective surface tension and bending rigidity of the elastic ring get modified under non-equilibrium growth conditions. Further, beyond a critical chemical potential driving force, the effective surface tension of the elastic ring is renormalized to zero and the elastic ring exhibits a buckling instability and undergoes a non-equilibrium morphological transformation (Fig.~\ref{fig:phases}). Such instabilities have been observed in experiments investigating the growth of model lipid membranes \cite{Solon2006} and can potentially have implications for biophysical processes such as membrane fission and endocytosis. We note that phenomenology similar to that described above can be observed in three-dimensional elastic membrane models (see SI Sec: 8~\cite{Supplementary}).

\begin{figure}[tbb]
\centering
\includegraphics[scale=0.38]{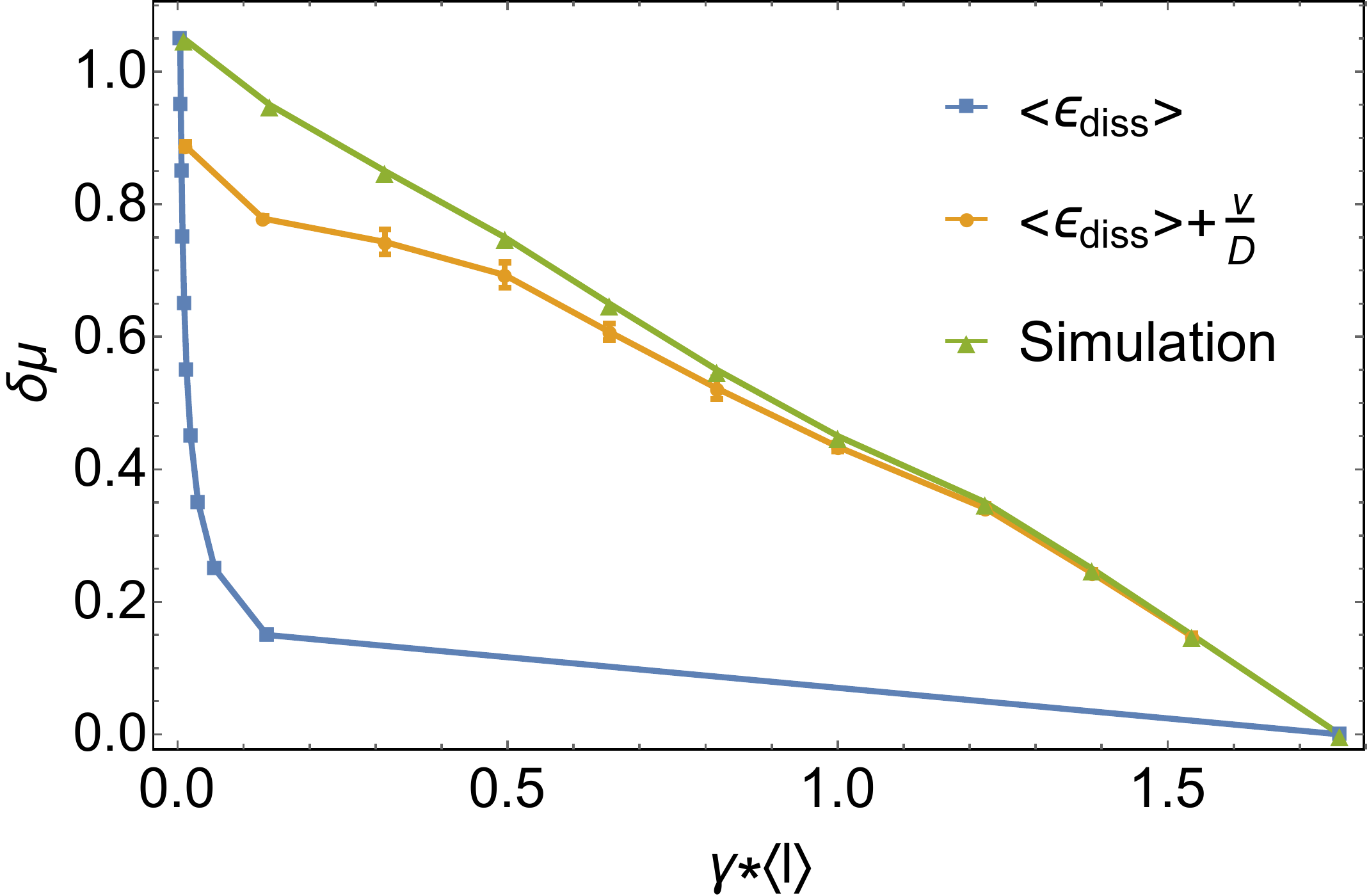}
\caption{Thermodynamic bounds on the surface tension as a function of the non-equilibrium driving force $\delta\mu$. The bound stipulated by the blue curve is from Eq.~\ref{eq:firstbound}. The bound stipulated by the orange curve is from Eq.~\ref{eq:secondbound}. The green curve is obtained by measuring $\gamma$ from simulations and is consistent with the bounds specified by Eq.~\ref{eq:firstbound} and Eq.~\ref{eq:secondbound}. These bounds provide rough estimates for the energetic costs required to modify the morphology and fluctuations in the elastic membrane. The error bars in the second bound is obtained from $95\%$ confidence intervals of fitting $\frac{v}{D}$.}\label{fig:PhaseDiagram}
\end{figure}

Using ideas from stochastic thermodynamics, we provide a thermodynamic prescription for how the surface tension and bending rigidity are modified by the non-equilibrium forces (Eqs.~\ref{eq:firstbound} and~\ref{eq:secondbound}). The thermodynamic prescription only requires information about the magnitude of the non-equilibrium chemical potential driving force, the equilibrium surface tension and bending rigidity, the average rate of growth, and fluctuations in the growth rate and is otherwise independent of the kinetic details of the growth process. The thermodynamic prescription is otherwise insensitive to any of the kinetic details used in the growth process and provides bounds on the energetic requirements to induce morphological transformations such as the above-described non-equilibrium buckling transition (Fig.~\ref{fig:PhaseDiagram}). 
Eq.~\ref{eq:secondbound} in particular is an adaptation of the recently derived thermodynamic uncertainty relations~\cite{Gingrich2016} to the problem of membrane growth.

A detailed proof for Eq.~\ref{eq:firstbound} and Eq.~\ref{eq:secondbound} is provided in SI Sec: 6~\cite{Supplementary}. This detailed proof shows how ideas like the thermodynamic uncertainty relation~\cite{Gingrich2016,Barato2015}\textendash these have typically been derived for Markov state models with a finite fixed number of states \textendash can be adapted and applied to non-equilibrium membrane growth problems. From this detailed proof, we also anticipate that the central thermodynamic result is not specific to the two-dimensional elastic membrane model and can be applied more broadly to study growth induced morphological transitions in three dimensional membranes (SI Sec: 8~\cite{Supplementary})~\cite{Mahadevan2019}. 
Together, our results form a set of design principles for controlling morphologies and material properties of membranes even in far from equilibrium conditions.


\begin{figure}[tbb]
\centering
\includegraphics[scale=0.29]{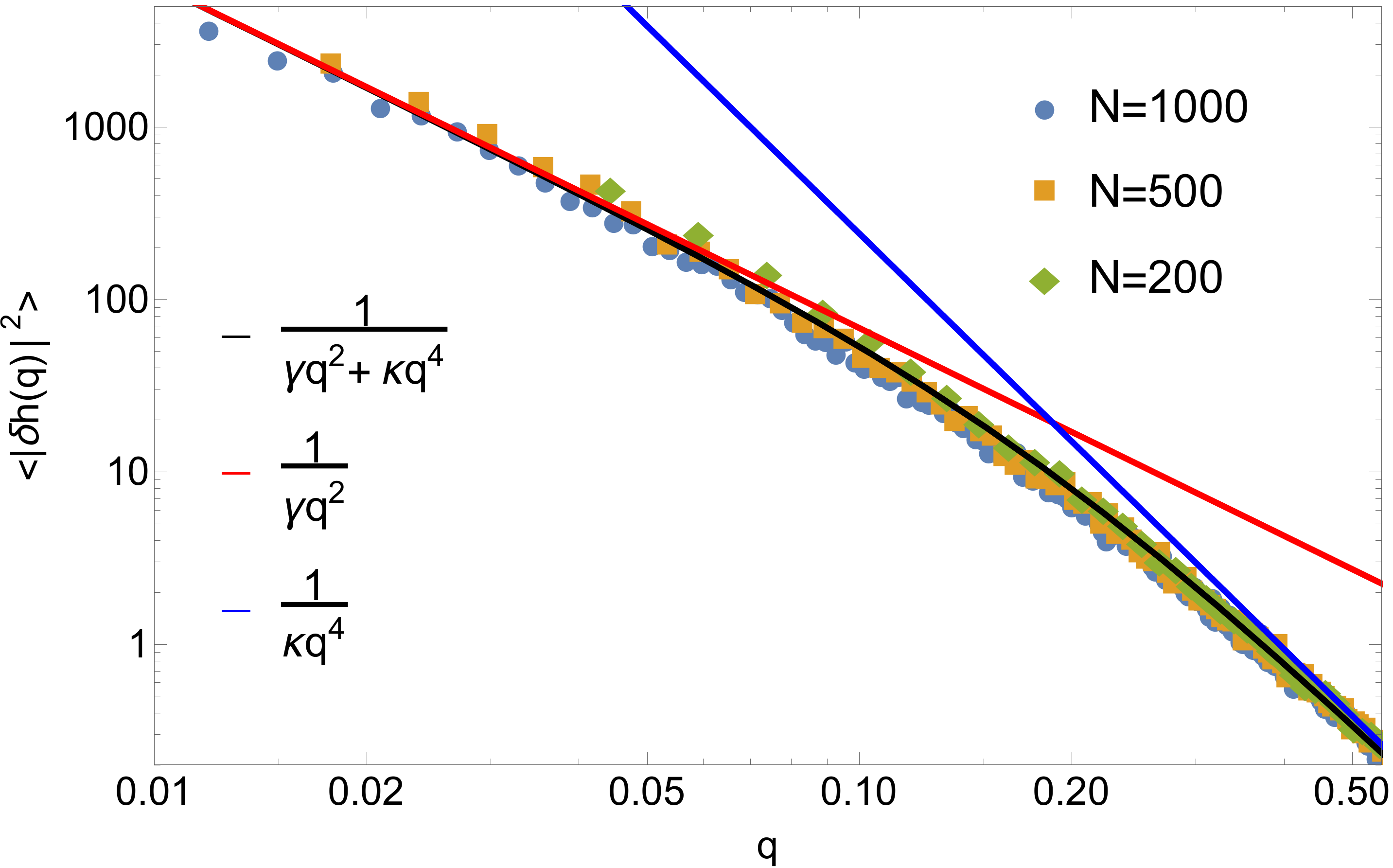}
\caption{Power spectrum of interfacial fluctuations at equilibrium. For small wavevectors, q, $|\delta h(q)|^2\propto q^{-2}$ while $|\delta h(q)|^2\propto q^{-4}$ at high q in agreement with expectations Eq.~\ref{eq:HelfrichFT}. The data here is for $k_s=4$ and $k_\theta = 6$. The diamond symbols are fluctuations of the assembly with 200 particles. The square symbols are fluctuations of the assembly with 500 particles. The circle symbols are fluctuation of the assembly with 1000 particles. Because the fluctuations here follow the Helfrich Hamiltonian, its standard deviation is exactly equal to its average magnitude due to the exponential nature of the distribution. Here $\gamma = 1.76$ and $\kappa =39.1$ from the fit} \label{fig:HelfrichScaling}
\end{figure}

\begin{figure}[tbb]
\centering
\includegraphics[scale=0.34]{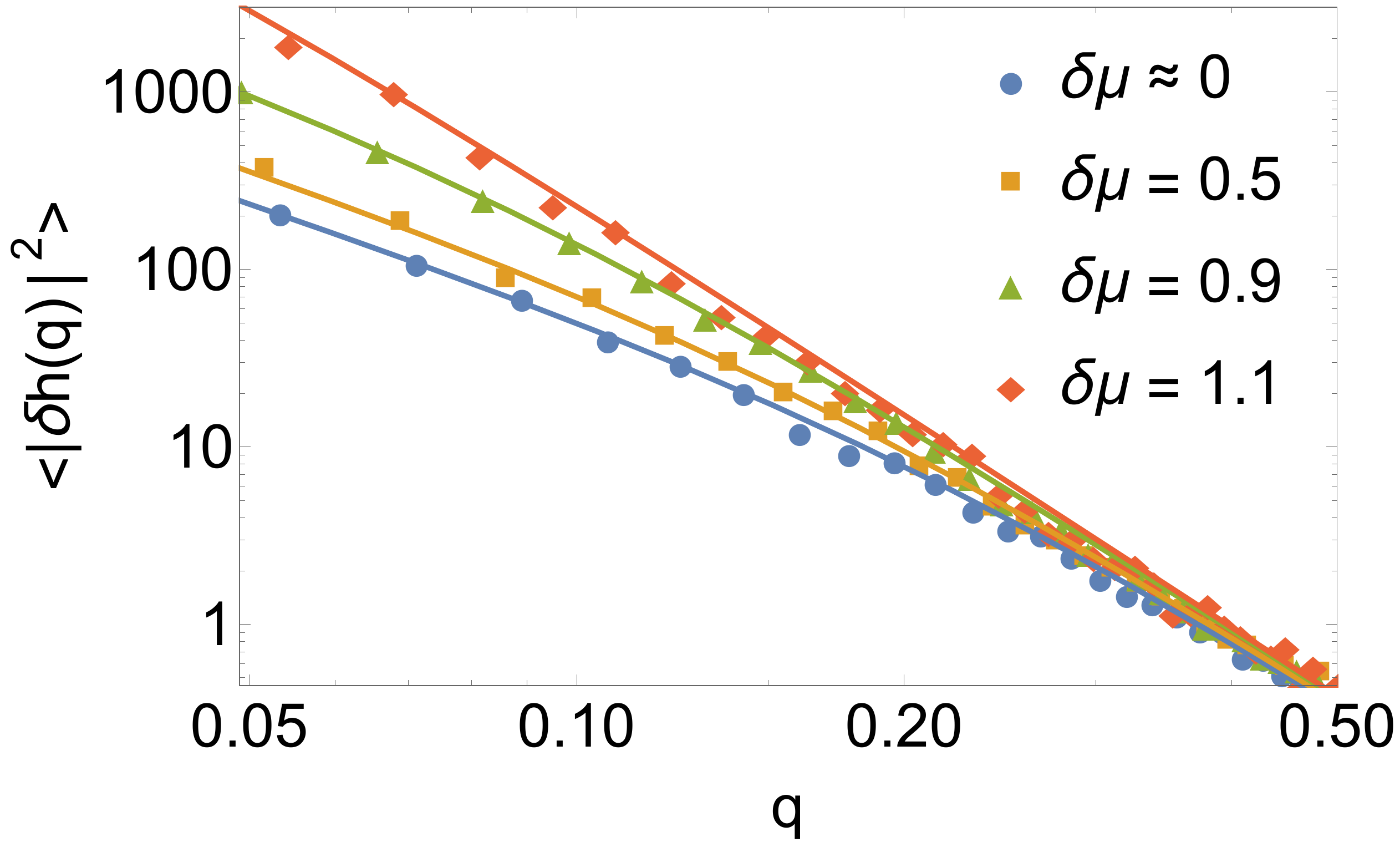}
\caption{Power spectrum of interfacial fluctuations at different $\delta\mu$. Fitting these curves to Eq.~\ref{eq:HelfrichFT} allows us to estimate $\gamma$ and $\kappa$ as described in the text. This analysis reveals that $\gamma_{\rm eff}$ decreases with increasing $\delta\mu$. The data here is for $k_s=4$ and $k_\theta = 6$.} \label{fig:Helfrich}
\end{figure}

\noindent{\it Simulations and results:} Our model consists system of particles in a ring like geometry interacting according to the Hamiltonian, 
\begin{equation}
\label{eq:Halmitonianl}
E=\sum_i^N{\frac{k_s}{2}(l_{i,i+1}-l_0)^2+k_\theta(\theta_i-\pi)^2}\,,
\end{equation}
where $l_{i,i+1}$ is the distance between particle $i$ and particle $i+1$, $l_0$ is the equilibrium distance and $\theta_i$ is the angle that particle $i$ makes with its neighbors. The growth dynamics of the Monte Carlo simulation are detailed in SI Fig S1. In short, in each Monte Carlo step, we attempt to add a particle from the bath or remove a random particle from the assembly with equal probability.   Events adding particles to the assembly are accepted with the probability $\rm{min}\{1,e^{(-\Delta E +\mu)/k_{\rm B} T}\}$ and events removing particles from the assembly are accepted with the probability $\rm{min}\{1,e^{(-\Delta E -\mu)/k_{\rm B} T}\}$. Here, the parameter $\mu$ can be regarded as the chemical potential of monomer units in the bath and $k_B T$ sets an energy scale. Unless specified otherwise, we set $k_{\rm B} T = 1$ for simplicity in the rest of the paper. 

The rate of growth of the elastic assembly can be tuned by varying the parameter $\mu$. Specifically, we find that there exists a \textit{coexistence} value of $\mu$, $\mu_{\rm coex}$, at which the assembly does not grow on average. The system is at equilibrium with its surroundings for this value of $\mu$. In the rest of the manuscript, we use the term equilibrium to refer to conditions where $\mu =\mu_{\rm coex}=\mu_{\rm eq}$, and the term non-equilibrium to refer to conditions where $\mu > \mu_{\rm eq}$. 

For values of $\mu$ above the coexistence value $\mu_{\rm eq}$, the system is driven away from equilibrium and the elastic ring polymer starts to grow. When $\delta\mu/k_{\rm B}T \equiv (\mu-\mu_{\rm eq})/k_{\rm B} T$ is small, the elastic assembly grows slowly and roughly retains its circular shape (see Movie: M1 in the SI). With increasing $\delta\mu/k_{\rm B} T$, the elastic assembly grows faster; its shape becomes more distorted (Movie: M2). Ultimately, the elastic assembly buckles resulting in spikes growing out of the circle as shown in Fig.~\ref{fig:phases} (Movie: M3). 

To study the above mentioned morphological changes (Fig.~\ref{fig:phases}), we examine how the fluctuations of the elastic ring polymer are modified as a function of its growth rate. Specifically, we divide the circumference of the assembly into $N$ equal segments with length $\langle l \rangle \equiv L/N$, where $N$ is the number of particles in the assembly at that instance of time and $L$ is the circumference of the assembly. 
 We then measure fluctuations in $\hat{h}(x_n)$ where $x_n\equiv n\langle l \rangle$, $\hat{h}(x_n)$ denotes the deviation of the $n^{\rm th}$ segment from the average radius of the elastic assembly, $\hat{h}(x_n)\equiv h(x_n) -\langle h \rangle$. The ensemble over which the fluctuations are measured was constructed by initiating simulations with a certain initial elastic assembly nucleus with size $N_0$ and allowing the nucleus to grow for a time $t_{\rm measure}$. In order to ensure that our results are not affected by choices of $N_0$ and $t_{\rm measure}$, simulations with multiple values of $N_0$ and $t_{\rm measure}$ were considered (See Fig.~\ref{fig:HelfrichScaling} and Fig. S7 and S8 in the SI~\cite{Supplementary}). In ensembles constructed in this manner, we measured $\langle |\delta h(q)|^2 \rangle$ where $\delta h(q)$ is the Fourier transform of the radial fluctuations defined with the convention: $\hat{h}(x_n)=\frac{1}{\sqrt{N}}\sum _q \delta h(q) \exp(iqx_n), q = \frac{2\pi m}{N\langle l \rangle}, m = 1, 2,...,N$. 

At or close to equilibrium, $\delta \mu /k_{\rm B} T \ll 1$, by measuring fluctuations and averaging over the above-described ensembles, we find that $\langle |\delta h(q)|^2 \rangle$ scales likes $q^{-2}$ in the low $q$ regime and scales like $q^{-4}$ in the high $q$ regime (Fig.~\ref{fig:HelfrichScaling}). This suggests that 
at equilibrium, the fluctuations of the elastic ring can be effectively described using the Helfrich Hamiltonian~\cite{W.Helfrich1973}:
\begin{equation}
E_{\rm eq}=\int \left \{\frac{\gamma_{\rm eq}}{2} (\nabla \hat{h})^2+  \frac{\kappa_{\rm eq}}{2} (\Delta \hat{h})^2\right \} dx.
\label{eq:Helfrich}
\end{equation}
Motivated by the scaling in Fig.~\ref{fig:HelfrichScaling}, we refer to the parameter $\gamma_{\rm eq}$ as an effective surface tension and the parameter $\kappa_{\rm eq}$ as an effective bending rigidity. We have tested the simulations with $k_s$ ranging from 2 to 4, and with $k_\theta$ ranging from 3 to 6. In these ranges, the fluctuations all follow the scaling of the Helfrich Hamiltonian. In addition, at equilibrium, $\gamma_{\rm eq}$ decreases with increasing $k_s$ and $k_{\theta}$. On the other hand, $\kappa_{\rm eq}$ seems to depend minimally on $k_s$ and decreases with decreasing $k_{\theta}$. We stress again that we are defining these elastic constants, $\gamma_{\rm eq}$ and $\kappa_{\rm eq}$ in the context of the ensembles defined above. 

\begin{figure}[tbb]
\centering
\includegraphics[scale=0.25]{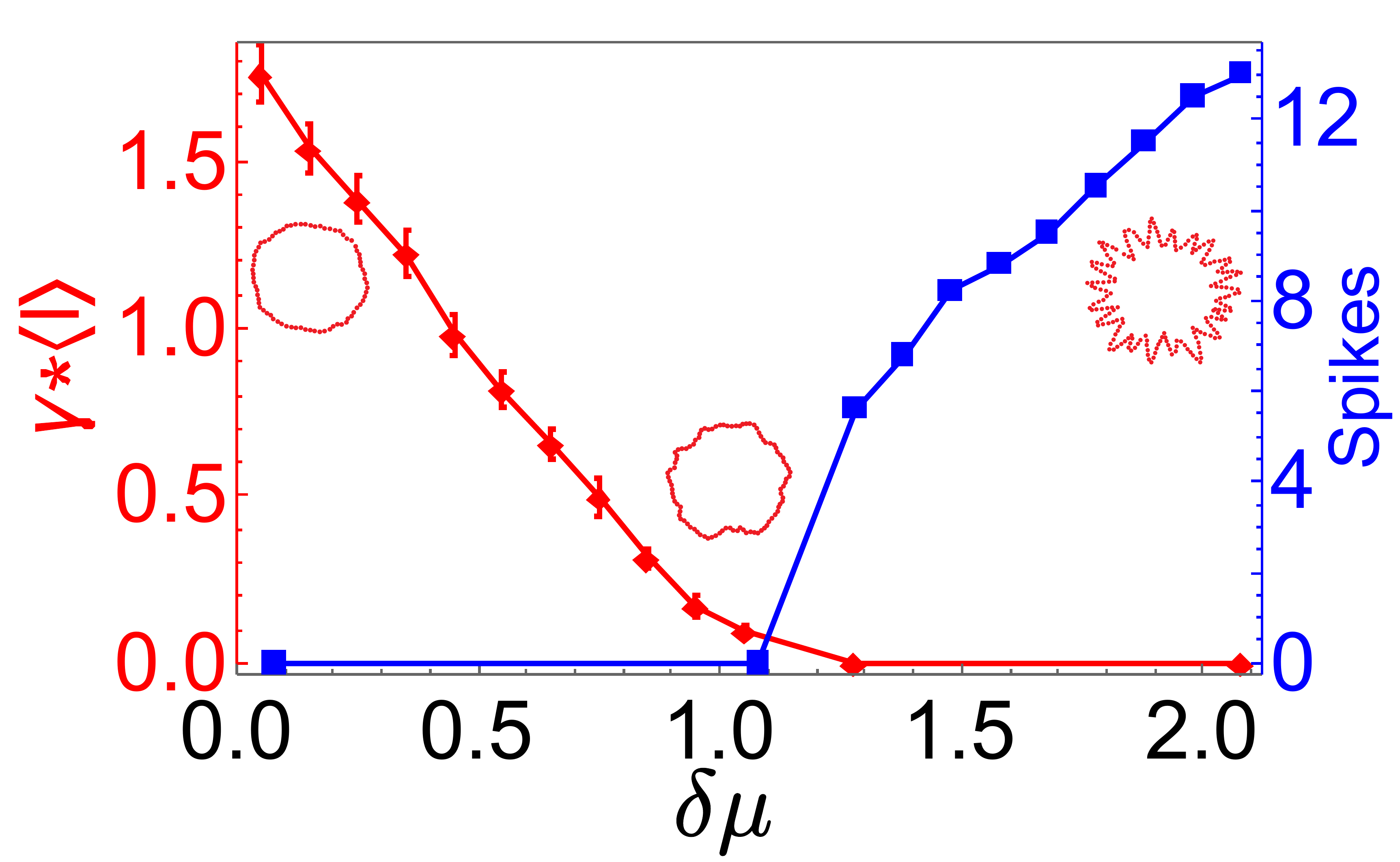}
\caption{Phase diagram for data at $k_s=4$ and $k_\theta = 6$. As $\delta\mu$ is increased, the effective surface tension $\gamma$ decreases eventually reaching $\gamma\approx 0$ for $\delta\mu \approx 1.1$.  Increasing $\delta \mu$ beyond this value induces a morphological change to a configuration with spikes. The data was obtained with $N_0=200$. The red curve in the figure represents the surface tension $\gamma$ of the assembly before the instability. The error bar represents the 95\% confidence interval from fitting. After the instability, $\gamma$ is negative and cannot be measure using the Fourier transform technique. We then use the number of spikes (the blue curve) in the assembly, which can be used to infer the instabilities' wavelength, to indicate the systems at different drive post the instability. The blue curve in the figure represents the number of spikes} \label{fig:HelfrichInstability}
\end{figure}

Even as the assembly starts to grow, Fig.~\ref{fig:Helfrich} shows that the radial fluctuations are still described by an effective Helfrich Hamiltonian with renormalized surface tension and bending rigidity, $\gamma$ and $\kappa$ respectively. Indeed, Fig.~\ref{fig:Helfrich} shows that the average $\langle |\delta h(q)|^2 \rangle$ is well described by 
\begin{equation}
\langle |\delta h(q)|^2 \rangle \propto \frac{k_B T}{(\gamma q^2 + \kappa q^4)}\,,
\label{eq:HelfrichFT}
\end{equation}
in accordance with Eq.~\ref{eq:Helfrich} with renormalized effective surface tension and bending rigidity values. Closer inspection of the effective surface tension and bending rigidity extracted from Fig.~\ref{fig:Helfrich} shows that the effective surface tension,$\gamma$ decreases as $\delta \mu$ is increased, dropping to $\gamma\approx 0$ at a critical value of $\delta \mu=\delta \mu_c$ (Fig.~\ref{fig:HelfrichInstability}). Beyond this point, the elastic assembly buckles and undergoes a morphological transformation to ring populated by \textit{}{spikes}. The number of spikes appearing in a process increases with $\delta\mu$ and is proportional to the initial size of the assembly (see Fig.~\ref{fig:HelfrichInstability}) and remains constant during the growing period. 
Reflecting the diminished effective surface tension cost under non-equilibrium conditions, the configurations with spikes allow the system to grow with minimal penalties for stretching. The bending rigidity does not seem to change by a large amount as indicated by fits obtained from Fig.~\ref{fig:Helfrich} (see Fig.~S7 in the SI~\cite{Supplementary}). In order to study how these elastic quantities depend on membrane size, we extracted the effective renormalized values of the surface tension and bending rigidity for multiple values of initial size $N_0$ and simulation time $t_{\rm measure}$. We find that to a good numerical approximation, the effective elastic constants, $\gamma$, $\kappa$ and values of the parameters $\mu_{\rm coex}$, and $\delta \mu_c$, are time and size independent in all our simulations as shown in Fig.~\ref{fig:HelfrichScaling} (see Fig.~S7 and S8 in the SI~\cite{Supplementary}). 


\noindent{\it A non-equilibrium thermodynamic theory for renormalization of surface tension and morphological changes:} We now use ideas from stochastic thermodynamics to understand the trade-offs between non-equilibrium driving (as characterized by $\delta \mu$) and morphological changes in the structure of this elastic ring system (as characterized by the renormalized constants $\gamma$ and $\kappa$). A detailed molecular derivation is provided in the SI (see SI Sec: 6)~\cite{Supplementary}. Here, we provide a phenomenological derivation. We begin by noting that our numerical results suggest that even when the elastic ring is not at equilibrium, its fluctuations can be described in terms of an effective energy landscape, (Fig.~\ref{fig:Helfrich}). In this case, using the principles of stochastic thermodynamics, an expression for the entropy of the growing elastic system can effectively be written down as~\cite{Nguyen2016,Esposito2012}
\begin{equation}
\label{eq:entropysystem}
 TS=\left\langle N \right\rangle_t \frac{-F_{\rm{eff}} + \langle E_{\rm{eff}} \rangle _N}{N}\,,
\end{equation}
where $E_{\rm{eff}}$ is the effective elastic energy of a configuration in terms of the renormalized material parameters $\gamma$ and $\kappa$, $F_{\rm{eff}}$ is the Helmholtz free energy appropriate to $E_{\rm{eff}}$, $\langle...\rangle_N$ is the average of all microscopic configurations of the assembly at size $N\gg 1$ and $\langle N \rangle_t$ is the average size of the elastic ring after it has been allowed to grow for a time $t$. Since $d\langle N \rangle_t/dt > 0$ under non-equilibrium conditions $\delta\mu >0$, the entropy of the system changes as a function of time. 

We can similarly compute the change in the entropy of the bath as it supplies monomers to the elastic assembly and maintains constant chemical potential conditions. Specifically, in the limit that the bath size is much larger than the size of any elastic assembly, the change in entropy of the bath after a time t, $\Delta S_{\rm bath}$, can be written as :
\begin{equation}
\label{eq:entropybath}
 T\Delta S_{\rm bath}=-\left\langle N \right\rangle_t \frac{-F_{\rm{eq}} + \langle E_{\rm{eq}} \rangle _N-N\delta\mu}{N}
\end{equation}
By combining Eq.~\ref{eq:entropysystem} and Eq.~\ref{eq:entropybath}, we can write down the total entropy of the process, which must be nonnegative according to the second law of thermodynamics~\cite{Esposito2012}:
\begin{equation}
\begin{split}
\label{eq:entropytotal}
\frac{dS_{\rm total}}{dt}&=\frac{dS}{dt} +\frac{dS_{\rm bath}}{dt}\\
&= \frac{d\langle N\rangle}{dt}\left ( \delta\mu-\langle\epsilon_{\rm diss}\rangle \right) \geq 0
\end{split}
\end{equation}
Here $\langle\epsilon_{\rm diss}\rangle=\left (\left\langle E_{\rm eq}-E_{\rm{eff}}\right\rangle _N-\left ( F_{\rm eq}-F_{\rm{eff}}\right )\right )/N$. $\langle\epsilon_{\rm diss}\rangle$ can be thought as the minimum work required to transform the energy landscape of the system from $E_{\rm eq}$ to $E_{\rm eff}$ using a driving force. The driving force here can come from many sources such as chemical activity and mechanical work. In this letter, the driving force here is the extra chemical potential we put into the bath.  For a growing system, $\frac{d\langle N \rangle}{dt}$ is positive thus reducing Eq.~\ref{eq:entropytotal} to:
\begin{equation}
\label{eq:firstbound}
\delta\mu-\langle\epsilon_{\rm diss}\rangle\geq0
\end{equation}

Tighter, and more informative bounds can be obtained by using the recently derived uncertainty relations that relate the entropy production to the fluctuations of various fluxes in the system~\cite{Barato2015, Gingrich2016,Nguyen2016}. For our purposes, we consider the fluctuations in the growth rate flux, $\dot{N}$. An application of the thermodynamic uncertainty relations then implies the following tighter bound: 
\begin{equation}
\label{eq:secondbound}
\delta\mu-\langle\epsilon_{\rm diss}\rangle\geq\frac{v k_{\rm B} T}{D}\, 
\end{equation}
where $v=\frac{d\langle N\rangle}{dt}$ is the growth rate of the assembly, and $D=\lim_{\tau\to\infty}\frac{\langle\Delta N^2\rangle}{2\tau}$  is the diffusion constant of the size fluctuations of the assembly. The equality in Eq.~\ref{eq:secondbound} is achieved in the linear response limit. 

The bounds in Eq.~\ref{eq:firstbound} and Eq.~\ref{eq:secondbound} constrain the allowed values of $\gamma$ and $\kappa$ given $\delta \mu$, the equilibrium elastic constants $\gamma_{\rm eq}, \kappa_{\rm eq}$ and the ratio $vk_{\rm B}T/D$. The non-equilibrium driving due to $\delta \mu$ can be used to maintain a growth rate and simultaneously renormalize fluctuations in the growing assembly. Eq.~\ref{eq:secondbound} assigns a thermodynamic cost for maintaining a growth rate, $v/D$, and a thermodynamic cost for renormalizing fluctuations, $\langle \epsilon_{\rm diss}\rangle$, and requires that net driving force, $\delta \mu$, be greater than the sum of the aforementioned thermodynamic costs. 

Note that Eqs.~\ref{eq:firstbound},~\ref{eq:secondbound} are minimally dependent on the kinetics of the growth process. A detailed microscopic proof of Eq.~\ref{eq:firstbound}, Eq.~\ref{eq:secondbound} is provided in the SI Sec 6~\cite{Supplementary}. This detailed microscopic proof does not require the membrane system to be constrained to two dimensions and can be readily applied to three dimensional membranes. Hence, we anticipate that the bounds in Eq.~\ref{eq:firstbound} and Eq.~\ref{eq:secondbound} can be applied to a broader class of non-equilibrium membrane growth processes (see SI Sec 6~\cite{Supplementary}). Finally we note that both the phenomenological derivation provided above and the detailed derivation provided in the SI assume that the fluctuations in the growth rate are minimally correlated with fluctuations in the configurations. As explained in the SI, formally this amounts to a mean field assumption that the growth rate of a particular configuration is simply proportional to the probability with which it is generated in the steady state. Extensions of Eqs.~\ref{eq:firstbound},~\ref{eq:secondbound} to non-mean field regimes are provided in the SI. Supported by numerical results detailed below, we note that the mean field assumption for growth rate statistics seems to work well for our membrane systems. In practice the mean field assumption can be shown to be exact for a class of lattice based non-equilibrium growth models~\cite{Nguyen2016}. 

\begin{figure}[tbb]
\includegraphics[scale=0.3]{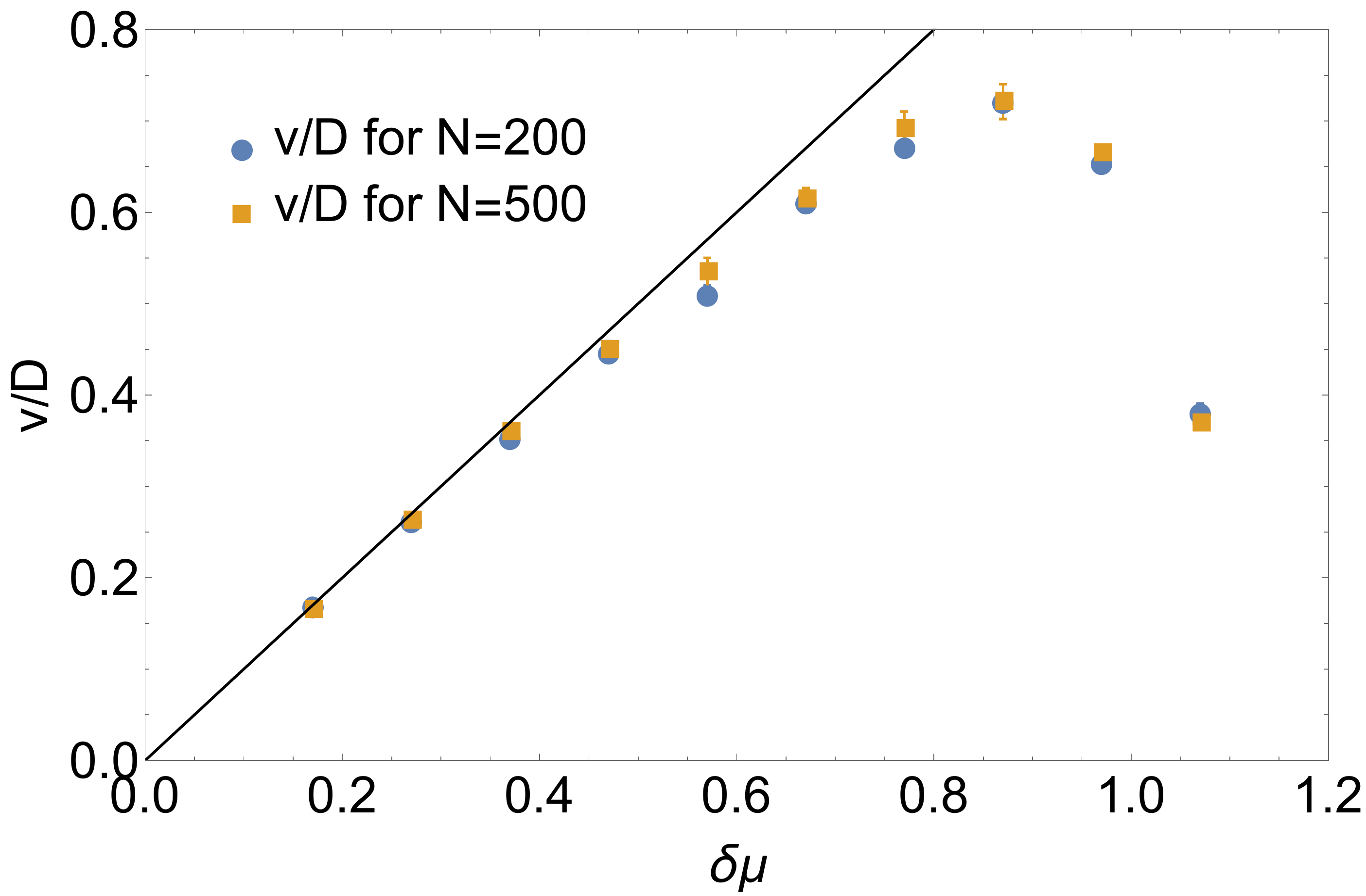}
\caption{$\frac{v}{D}$ vs. $\delta\mu$. In this case, $k_{\rm B}T$ is set to 1. The dark line is predicted from linear response. The error bar represents the 95\% confidence interval from the fit $\delta\mu=v/D$. The blue dot is from the assemblies of 200 particles, while the orange dot is from the assemblies of 500 particles. The two measurements overlaps with some minor error.}
\label{fig:LinearResponse}
\end{figure}

Before proceeding to use Eq.~\ref{eq:secondbound} to elucidate how the thermodynamic driving forces control the renormalization of material properties and morphologies, we first consider Eq.~\ref{eq:secondbound} without the term non-negative term $\langle \epsilon_{\rm{diss}}\rangle$~\footnote{The statement $\langle\epsilon_{\rm diss}\rangle\geq 0$ can be proven by applying Jensen's inequality to: $\langle\exp\left[-(E_{\rm{eq}}-E_{\rm{eff}})\right]\rangle_N=\exp\left[-(G_{\rm{eq}}-G_{\rm{eff}})\right]$}. The bound in Eq.~\ref{eq:secondbound} reduces to the following relation between driving force $\delta \mu$ and ratio $v/D$, $\delta\mu\geq\frac{v k_B T}{D}$. In Fig.~\ref{fig:LinearResponse} we numerically verify that our simulations at two different sizes do indeed satisfy this simplified connection. Further, in the limit of slow driving, $\delta \mu/k_B T\ll 1$, Fig.~\ref{fig:LinearResponse} reveals that most of the driving force is used up in maintaining a growth rate with very little remaining for renormalization of material parameters. In this limit, $\delta \mu \approx v/D$. 
At larger values of the driving, $\delta \mu$ deviates significantly from $v/D$. Larger value of the thermodynamic cost associated with renormalized fluctuations, $\langle \epsilon_{\rm diss}\rangle$, are hence allowed by our thermodynamic bound in these regimes. Indeed, our simulations (Fig.~\ref{fig:phases}) show how a dramatic change in morphologies can be achieved for $\delta \mu/k_B T \approx 1$. 

We will now use Eq.~\ref{eq:secondbound} to understand how $\gamma$ can be controlled by tuning $\delta \mu$. As first approximation, given the relatively slow renormalization of the bending rigidity, we will set $\kappa=\kappa_{\rm eq}$ (see Fig S7 in SI). Within this approximation (see SI Sec 7~\cite{Supplementary}), we obtain the following simplified expression for $\langle \epsilon_{\rm diss} \rangle$: 
\begin{equation}
\label{eq:epsilonexpress}
 \langle\epsilon_{\rm diss}\rangle =\frac{k_{\rm B}T\lambda}{4\pi}\left ( \frac{(\gamma + \gamma_{\rm eq})\phi-2\sqrt{\gamma\gamma_{\rm eq}}\phi^*}{\sqrt{\gamma \kappa_{\rm eq}}}-\frac{2\pi \xi}{\lambda} \right)
\end{equation}
Here $\phi=\arctan(\frac{2\pi\sqrt{\kappa_{\rm eq}}}{\lambda\sqrt{\gamma}}), \phi^*=\arctan(\frac{2\pi\sqrt{\kappa_{\rm eq}}}{\lambda\sqrt{\gamma_{\rm{eq}}}})$, $\xi=\ln(\frac{\gamma_{\rm{eq}}\lambda^2+4\pi^2\kappa_{\rm eq}}{\gamma\lambda^2+4\pi^2\kappa_{\rm eq}})$, and $\lambda$ is the smallest wavelength allowed by the assembly which we will take to be $l_0$. Using this expression for $\epsilon_{\rm diss}$, Eq.~\ref{eq:firstbound} and Eq.~\ref{eq:secondbound} can be used to predict bounds on how $\gamma$ changes with the non-equilibrium driving $\delta \mu$. These predictions are plotted in Fig.~\ref{fig:PhaseDiagram} alongside the scaling of $\gamma$ with $\delta \mu$ extracted from simulations. 

Fig.~\ref{fig:PhaseDiagram} shows how ideas from stochastic thermodynamics can be used to predict how material properties such as the surface tension can be modified in the presence of non-equilibrium forces. In particular, the lower bound suggested by Eq.~\ref{eq:secondbound} is surpisingly close to the actual non-equilibrium driving force $\delta \mu$ required to renormalize membrane tension and induce morphological transformations ($\gamma\sim 0$). Unlike the usual approaches, the bounds here do not require extensive knowledge of the kinetics of the system. Indeed, as evidenced by the performance of the bound in Eq.~\ref{eq:secondbound} in Fig.~\ref{fig:PhaseDiagram}, our results show how a large component of the non-equilibrium renormalization of membrane material properties is effectively controlled by two (experimentally accessible) parameters, the driving force $\delta \mu$, and the ratio $v/D$. 




The role played by non-equilibrium forces in biological processes such as those responsible for modulating cell shapes and dynamics is well established~\cite{McMahon2005,Stachowiak2012,Chen2016,Turlier2016,Rao2001,Solon2006}. In this paper we have shown how ideas from stochastic thermodynamics, in particular an adaptation of the recently derived thermodynamic uncertainty relations, can be used to obtain general non-equilibrium thermodynamic constraints on membrane morphologies and material properties. Our thermodynamic bounds are minimal dependent on the details of the kinetic processes responsible for membrane growth. We anticipate that such thermodynamic ideas will find broad applicability and reveal how material properties and morphologies can be robustly controlled even far from equilibrium. 

The authors acknowledge support from NSF DMR-
MRSEC  1420709, NSF GFRP and the  University  of  Chicago
 
%

\end{document}